\documentclass{article}

\usepackage{booktabs}
\usepackage{graphics}
\usepackage{graphicx}
\usepackage{listings}
\usepackage{nameref}
\usepackage{algorithm}
\usepackage{algpseudocode}
\algrenewcommand\algorithmicrequire{\textbf{Input:}}
\algrenewcommand\algorithmicensure{\textbf{Output:}}
\usepackage{bm}
\usepackage{amsmath}
\usepackage{amssymb}
\usepackage{caption} 
\usepackage{multirow}

\usepackage{arxiv}

\usepackage[utf8]{inputenc} % allow utf-8 input
\usepackage[T1]{fontenc}    % use 8-bit T1 fonts
\usepackage{hyperref}       % hyperlinks
\usepackage{url}            % simple URL typesetting
\usepackage{booktabs}       % professional-quality tables
\usepackage{amsfonts}       % blackboard math symbols
\usepackage{nicefrac}       % compact symbols for 1/2, etc.
\usepackage{microtype}      % microtypography
\usepackage{lipsum}
\usepackage{graphicx}
\graphicspath{ {./images/} }

\title{Phage family classification under \textit{Caudoviricetes}: a review of current tools using the latest ICTV classification framework}

\author{
  Yilin Zhu \\
  Dept. of Electrical Engineering\\
  City University of Hong Kong\\
  Kowloon, Hong Kong SAR, China\\
  \texttt{zyl666777@gmail.com} \\
  \And
 Jiayu Shang \\
  Dept. of Electrical Engineering\\
  City University of Hong Kong\\
  Kowloon, Hong Kong SAR, China\\
  \texttt{jyshang2-c@my.cityu.edu.hk} \\
  \And
  Cheng Peng \\
  Dept. of Electrical Engineering\\
  City University of Hong Kong\\
  Kowloon, Hong Kong SAR, China\\
  \texttt{cpeng29-c@my.cityu.edu.hk} \\
  \And
 Yanni Sun \\
  Dept. of Electrical Engineering\\
  City University of Hong Kong\\
  Kowloon, Hong Kong SAR, China\\
  \texttt{yannisun@cityu.edu.hk} \\
}

\begin{document}

\maketitle
\begin{abstract}
Bacteriophages, which are viruses infecting bacteria, are the most ubiquitous and diverse entities in the biosphere. There is accumulating evidence revealing their important roles in shaping the structure of various microbiomes. Thanks to (viral) metagenomic sequencing, a large number of new bacteriophages have been discovered. However, lacking a standard and automatic virus classification pipeline, the taxonomic characterization of new viruses seriously lag behind the sequencing efforts.  
In particular, according to the latest version of ICTV, several large phage families in the previous classification system are removed. Therefore, a comprehensive review and comparison of taxonomic classification tools under the new standard are needed to establish the state-of-the-art. In this work, we retrained and tested four recently published tools on newly labeled databases. We demonstrated their utilities and tested them on multiple datasets, including the RefSeq, short contigs, simulated metagenomic datasets, and low-similarity datasets. This study provides a comprehensive review of phage family classification in different scenarios and a practical guidance for choosing appropriate taxonomic classification pipelines. To our best knowledge, this is the first review conducted under the new ICTV classification framework. The results show that the new family classification framework overall leads to better conserved groups and thus makes family-level classification more feasible. \\
\textbf{Contact:} \href{yannisun@cityu.edu.hk}{yannisun@cityu.edu.hk}\\
\end{abstract}


\newpage

\section{Introduction}

% concept & variety
Bacteriophages (aka phages) are viruses that infect bacteria \cite{mcgrath2007bacteriophage}. Phages are the most abundant biological entities on Earth. It is estimated that there are more than $10^{31}$ bacteriophages on the planet, outnumbering every other organism on Earth combined \cite{suttle2005viruses, lafee2017novel}. 
In most microbial communities, phages play a crucial role by shaping and maintaining microbial ecology \cite{koskella2013understanding, thingstad2000elements}, facilitating co-evolutionary relationships \cite{cobian2016viruses, hoyles2014characterization, silveira2016piggyback}, and promoting microbial evolution through horizontal gene transfer \cite{brown2015transfer, chiang2019genetic}. 

Phages are diverse in size, morphology, and genomic organization \cite{chow2015biogeography,ackermann2006classification}. They have a variety of structural morphologies, among which tailed double-stranded DNA (dsDNA) phages \cite{kauffman2018major, brum2013global} are the most abundant. Besides dsDNA phages, there are also phages with single-stranded DNA (ssDNA) \cite{lim2015early}, single-stranded RNA (ssRNA) \cite{loeb1961bacteriophage} or double-stranded RNA (dsRNA) \cite{mertens2004dsrna}. Phages also have a wide range of genome sizes. Recently, an increasing number of megaphages (\textgreater 200kbp) have been sequenced, demonstrating unique genomic features \cite{yuan2017jumbo}. Because of the high diversity of genomes, phages infecting different hosts typically have a low similarity. However, phages that infect the same host may also have considerable differences in their genomes \cite{hatfull2008bacteriophage,krupovic2011genomics}.

It is now demonstrated that phages can be found in a wide variety of environments, including aquatic ecosystems \cite{paul2002marine, guttman2005basic}, human gut \cite{sutton2019gut, manrique2017human}, and soil \cite{williamson2017viruses, chow2015biogeography}. The first viral metagenome of uncultured marine viral communities was published in 2002 \cite{breitbart2002genomic}. Phages can shape the composition and function of underlying ecosystems through two different lifestyles: temperate and virulent. Temperate phages will integrate their genomes into bacterial chromosomes and replicate with their host. They will maintain this state, which is also called prophages, until being induced by the environment's condition, such as appropriate temperature and pH value. Then, temperate phages will enter the lytic cycle to kill the host \cite{campbell2003future,howard2017lysogeny}. In contrast, virulent phages do not integrate their genomes into the hosts. They stay in the lytic cycle and kill the hosts after replicating themselves \cite{hobbs2016diversity}. 

% importance & understanding limitation
The unique properties and life styles make phages key players in multiple applications. For example, phage therapy is a promising strategy for treating bacterial infections, particularly those with antibiotic-resistant bacteria. It has been found that intravenous phage preparations could treat \textit{Staphylococcus aureus} that induced pneumonia in mice \cite{saussereau2012bacteriophages, oduor2016experimental}. In addition, phages can be used to treat gastrointestinal infections. It has been demonstrated that phages are effective in reducing intestinal pathogens and have less impact on the composition of the intestinal microbiota compared to antibiotics \cite{galtier2016bacteriophages,nale2016bacteriophage, jaiswal2013efficacy, gutierrez2020phage}. Moreover, phages are important in food safety. The use of specific phage treatments in the food industry can prevent product spoilage and limit the spread of bacteria, providing a safe environment for animal and plant food production \cite{sillankorva2012bacteriophages, garcia2008bacteriophages, coffey2010phage, gutierrez2017applicability}. 

However, despite the abundance and importance of phages in various ecosystems, our understanding of phages is still very limited. According to the database supported by the National Center for Biotechnology Information (NCBI), the number of identified phages in class \textit{Caudoviricetes} changed from 1,359 in 2015 to 4,483 in 2022 in the RefSeq database, which is tripled in size. Besides the reference genomes, there are roughly 63,588 assembled phages belonging to Class \textit{Caudoviricetes} in the Genbank database in 2022, an almost five fold increase compared to 2015 (16,232). However, the characterization of phages cannot keep pace with the fast increase of the sequencing data.

% ICTV -> ICTV limitations
Assigning phages into different taxonomic groups is a fundamental step following phage discovery. The official taxonomy was established by the International Committee on Taxonomy of Viruses (ICTV) \cite{adams201750}, which organizes viruses in several taxonomic levels, including class, order, family, subfamily, genus and so on. Within the ICTV, the Bacterial and Archaeal Viruses Subcommittee (BAVS) is responsible for the phages' taxa. BAVS classifies phages based on a variety of phage properties, including the molecular composition of the genome (ss/ds, DNA, or RNA), the morphology, the structure of the capsid, and the host range \cite{dion2020phage}. Recently, with the increasing availability of viral genomes, using genomes for taxonomic classification has become more widely accepted \cite{lefkowitz2018virus}. Due to the extensive sequencing efforts for virus discovery, ICTV cannot catch up with the sheer number of newly identified phages, and thus many viruses are still not classified. One challenge behind this delay is the lack of standard, accurate, and comprehensive taxonomic classification tools for phages. Indeed, phage classification is not a trivial problem.  The taxonomic standard in ICTV is constantly changing as new phages are discovered. Recently, ICTV updated the phage classification system in August 2022, in which several major families in the previous ICTV system are removed, such as \textit{Siphoviridae}, \textit{Podoviridae}, and \textit{Myoviridae}. These changes can significantly affect the performance of family classification. To our best knowledge, no quantitative evaluations of the performance change have been conducted. Table \ref{tab:similarity} shows the average similarity (calculated by Dashing \cite{baker2019dashing}) of the largest four families in the old and new ICTV taxonomy classification systems. The updated families are more conserved as shown by the increased average similarity, making family-level classification more feasible. 

\begin{table*}[!h]
\centering
\renewcommand\arraystretch{1.3}
\caption{The average pairwise Dashing similarity of the four largest phage families under \textit{Caudoviricetes}}
\begin{tabular}{ p{3.7cm}p{3.7cm}p{3.7cm}p{3.7cm} }
\hline
\multicolumn{2}{c}{\textbf{Old version ICTV}}&\multicolumn{2}{c}{\textbf{New version ICTV}}\\
\hline
\textbf{Phage Family}  & \textbf{Similarity} & \textbf{Phage Family}  & \textbf{Similarity}\\
\hline
\textit{Siphoviridae}        & 0.0129 & \textit{Autographiviridae}  & 0.0171 \\
\textit{Myoviridae} & 0.0157  & \textit{Straboviridae} & 0.0748  \\
\textit{Autographiviridae}  & 0.0171 & \textit{Herelleviridae} & 0.0519 \\
\textit{Podoviridae}  & 0.0206   & \textit{Drexlerviridae} & 0.0432 \\
\hline
\end{tabular} 
\label{tab:similarity}
\end{table*}

%Tools development
Available taxonomic classification tools often have different designs and were tested on different datasets by their authors. Without a comprehensive comparison on the same training/reference data set and test set, it is difficult for users to choose the most appropriate solution for their needs. This paper presents a comprehensive benchmark of the main players in phage taxonomic classification under the latest ICTV standard. The remaining of this review is organized as follows. First, we will describe the main methods/models for existing phage taxonomic classification approaches and discuss whether they can be retrained/used under the new ICTV taxonomy standard. Then, we evaluate the four representative approaches that can be retrained by newly labeled sequences in different usage scenarios. In particular, we tested these tools on complete virus genomes, short contigs, simulated metagenomic datasets, and low-similarity datasets. In addition, we conducted a leave-one-family-out experiment to test whether these tools can recognize out-of-distribution sequences. By comparing their performance and analyzing the underlying reasons, we draw conclusions and provide guidance for users about choosing the most appropriate tools for different scenarios.

\section{Approaches for phage taxonomic classification}

Most phage taxonomic classification approaches can output classification results in different ranks, such as order, family, and genus. In this review, we focus on comparing different tools' performance at the family level because of the following reasons. First, the taxonomy by ICTV is under constant changes, which affects the total genus number significantly. For example, there are 735 genera in the ICTV database released in 2016. However, the number of genera increased to 2,224 in 2020. The overhaul of the genus-level taxonomy can make the definition of ``ground truth'' ambiguous. In addition, hundreds of rare genera only contain one phage, making the construction of reference and test set difficult. Second, classification at higher taxonomic ranks is usually easier than at lower ranks due to the smaller inter-class similarities and more abundant sequences in each class. Thus classification at orders or above is not as challenging as family classification. \textit{Caudoviricetes}, a class of phage known as the tailed phages whose hosts are phage and archaea, contains the majority of the total phage sequences and can be classified by almost all of the tools mentioned above, we thus focus on the classification of the families under \textit{Caudoviricetes} in this work.

\begin{table*}[]
\normalsize
\centering
\vspace{-0.5cm}
\caption{Overview of bioinformatic approaches used for phage taxonomic classification.}
\renewcommand\arraystretch{2}
{
\begin{tabular}{p{2cm}p{1.5cm}p{8.5cm}p{1.2cm}p{1cm}}
\hline
\textbf{Name}                   
& \textbf{Year} & \textbf{Description} & \textbf{Input data}  & \textbf{Lowest level}\\ \hline
Phage Proteomic Tree \cite{rohwer2002phage} & 2002 
& It uses the BLASTP distance and protein distance scores (similarity between two proteins) to generate phage proteomic trees, which can describe the relationships between different phages and can serve as a genome-based classification system for phages.
& protein sequences & Family\\ 

POGs \cite{kristensen2013orthologous} & 2013    
& It provides a collection of orthologous genes clusters from phages, represented as profiles. It extracts virus-specific genes, and then classifies phages by aligning query sequences against the marker genes utilizing BLAST. 
& genome sequences & Genus \\

GRAViTy \cite{aiewsakun2018genomic}& 2018 
& It conducts taxonomic classification by computing sequence relatedness between viruses using Composite Generalized Jaccard (CGJ) distances that integrate homology detection outputs and shared genomic features.
& genome sequences & Genus\\

CCP77 \cite{low2019evaluation}& 2019 
& A Phylogeny-based taxonomic classification for \textit{Caudovirales}, inferring a concatenated Caudovirales protein (CCP77) tree based on the concatenation of protein markers using a maximum-likelihood method.
& genome sequences & Genus \\

ClassiPhage \cite{chibani2019classifying,Chibani558171}& 2019
& It uses a set of phage-specific Hidden Markov Models (HMMs) generated from clusters of related proteins for phage taxonomic classification. Classiphage 2.0 adds an Artificial Neutral Network (ANN) in the models.
& protein sequences & Family\\

vConTACT \cite{bolduc2017vcontact,bin2019taxonomic}& 2019 
& 
A network-based application utilizing whole genome gene-sharing profiles, which integrates distance-based hierarchical clustering and confidence scores for virus classification.
& protein sequences & Genus \\

CAT \cite{von2019robust}& 2019 
& It can provide taxonomic classification for contigs or contig bins utilizing DIAMOND BLASTP homology search for open reading frames (ORFs).
& genome sequences & Species
 \\
MMseqs2 \cite{mirdita2021fast}& 2021
& 
A protein-search-based taxonomy classification tool for contigs. It assigns taxa for each possible protein product and uses weighted voting to assign taxonomic labels.
& genome sequences & Species\\
 
VPF-Class \cite{pons2021vpf}& 2021
& It automates the classification by assigning the proteins to a set of Viral Protein Families (VPFs), which are then used to estimate the similarity between query genomes with classified genomes.
& genome sequences & Genus
\\
PhaGCN \cite{shang2021bacteriophage}& 2021 
& A semi-supervised learning model. It formulates the taxonomic classification problem as a node classification problem in a knowledge network, which is constructed by combining the DNA sequence features and protein sequence similarity. 
& genome sequences & Family

\\\hline
\end{tabular}}
\label{tab:tool-compare}
\vspace{0.5cm}
\end{table*}

The phage taxonomic classification methods are summarized in Table \ref{tab:tool-compare} following the chronological order, which includes a brief description, publication year, required input data type, and the lowest predicted level of each tool. 
A majority of these tools conduct phage taxonomic classification based on sequence comparison, utilizing nucleotide-level or protein-level similarity between a query virus and the reference database. 
The comparison-based methods differ in their constructed reference database, the alignment method, and how they utilize these alignments. Both pairwise sequence alignment and hidden Markov model (HMM)-based profile alignments are commonly used. Multiple tools construct virus protein families and use them as marker genes. Using markers usually incurs less memory usage than using all phage genomes. But newly sequenced phages with novel genes may not be aligned to any marker gene families and thus cannot be assigned to a known class. 
Learning-based models have also been applied to phage classification. Learning models can automatically infer the sequence patterns in phage genomes of different families and use the learned features for automatic classification. A more detailed description of these tools is provided below.

Phage Proteomic Tree \cite{rohwer2002phage,nishimura2017viptree} is a relatively early program providing phage genome classification down to the family level. It extracts protein sequences from virus genomes and clusters these sequences using BLASTP \cite{altschul1997gapped}. Then the clusters in Phage Proteomic Tree are refined and scored. Finally, the alignment scores are converted to distances, which were used to generate the final tree using the neighbor-joining algorithm.

Taxon-specific signature genes can be identified in most virus taxa. POGs (Phage Orthologous Groups) \cite{kristensen2013orthologous} is a collection of clusters of orthologous genes from phages, presented as profiles (multiple sequence alignment). The viral families of POGs are filtered as \textit{`Viruses[Organism] NOT cellular organisms [ORGN] NOT srcdb\_refseq[PROP] AND vhost bacteria[filter] AND “complete genome”[All Fields]} in NCBI. Signatures are extracted for each taxon, and we can use BLASTP to search for matches among the viral protein sequences. POGs are designed to be well suited for defining taxon-specific signature genes, and the profiles built from POGs are more sensitive and specific to search for signature genes in a given dataset.

GRAViTy \cite{aiewsakun2018genomic} also extracts protein sequences from virus genomes and cluster these sequences using BLASTP \cite{altschul1997gapped}. GRAViTy generates protein profile hidden Markov models (PPHMMs) and genomic organization models (GOMs) based on the sequences from BLASTP-based clustering. Then it computes Composite Generalized Jaccard (CGJ) similarity scores (a geometric mean of the two generalized Jaccard scores computed for a pair of PPHMM signatures and a pair of GOM signatures) between each sequence pair to construct the heat map and dendrogram and estimate sequences' relatedness. GRAViTy requires users to choose reference database freely but need sequences in GenBank format as input.

CCP77 \cite{low2019evaluation} applies a concatenated protein phylogeny for the classification of tailed dsDNA viruses belonging to the specific order \textit{Caudovirales}. Classiphage \cite{chibani2019classifying,Chibani558171} uses phage-specific Hidden Markov Models (HMMs) \cite{eddy2011accelerated} profiles generated from clusters of related proteins for classification. The HMM profiles are built using the produced multi-sequence alignment files by the ``hmmbuild” command. Classiphage 2.0 additionally trains an Artificial Neutral Network (ANN) using phage family-proteome to phage-derived HMMs scoring matrix, which can classify more phage families and include more features than its previous version. 

vConTACT \cite{bolduc2017vcontact,bin2019taxonomic} is a high-throughput network-based approach utilizing whole-genome gene-sharing profiles. It clusters the input viral genomes together with characterized genomes. The genomes in the same cluster indicate the same family or genus, and the predicted family can be inferred if there are characterized genomes in the same cluster.

CAT \cite{von2019robust} provides taxonomic classification using homology searches. It uses DIAMOND BLASTP to identify homologous sequences and then assigns query sequences into taxa with a voting approach. The authors of CAT show that using the best hit strategy can lead to low specificity and thus design a more robust strategy based on multiple hits. Users can select the reference database and tune the setting, which is more flexible than some other tools. Moreover, it has a very low memory usage. 

MMseqs2 \cite{mirdita2021fast} is a fast contig taxonomic assignment tool. Similar to CAT, it conducts protein homology search against reference sequences and uses majority vote to assign the most specific taxon for a contig. With some optimizations and adoption of 2bLCA \cite{hingamp2013exploring}, MMseqs2 circumvents the need of adjusting a parameter in CAT and achieves faster speed on the tested bacterial and eukaryotic datasets.  It allows users to supply a customized reference database.

VPF-Class \cite{pons2021vpf} provides both taxonomic classification and host prediction for input viral genomes. It compares predicted proteins against the set of constructed Viral Protein Families (VPFs) (from the IMG/VR system). Then it derives taxonomic classifications and confidence scores from the list of VPFs detected on each query genome. However, VPF-Class does not require users to download and select the reference datasets. 

PhaGCN \cite{shang2021bacteriophage} is a semi-supervised learning model for phage taxonomic classification developed by our team. This model constructs a knowledge graph by combining the DNA sequence features learned by Convolutional Neural Networks (CNN) and protein sequence similarity gained from the gene-sharing network. The learning model can incorporate the automatically learned features for each contig. However, unlike sequence comparison-based approaches, PhaGCN only accepts phage-like sequences as input. Thus, a pre-processing step is needed for detecting those contigs from metagenomic data. A number of tools, such as VirFinder \cite{ren2020identifying}, Seeker \cite{auslander2020seeker}, and PhaMer \cite{Shang_2022} can be applied in the pre-processing step.

\section{Experiments and Results}

Because of the changes in the ICTV classification system, the models/reference databases need to be updated using the latest labeled sequences. However, not all the tools in Table \ref{tab:tool-compare} can be updated easily. Among them, only CAT, GRAViTy, PhaGCN, MMseqs2, and vConTACT 2.0 allow users to change their reference databases or retrain the models with reasonable efforts. The others do not specify the feasibility of changing models or reference databases in the descriptions. The source code of CCP77 is only available on request but not to the public. The code of GRAViTy released at GitHub is the alpha version and the author mentioned that they are currently working on a new and improved version that is more user-friendly and written in python3. Nevertheless, we downloaded and installed the alpha version of GRAViTy. The alpha version is computationally expensive and requires 30 hours to build a reference database with about 1200 genomes and another 25 hours to process just 300 queries. Therefore, we focus on evaluating the performance of the four tools: PhaGCN, vConTACT 2.0, CAT, and MMseqs2. These tools were recently published and demonstrated good performance in their own or others' tests. In addition, the corresponding codes and tools are still under maintenance. None of them requires an internet connection or a web server. To mimic the scenario of applying these tools to datasets without known taxonomic composition, we apply all these tools with their default parameters, which are optimized by the authors.
The commands for running all these tools are available in the Supplementary File.
All the tools were run on IntelVR$^\circledR$XeonVR$^\circledR$ Gold 6258 R CPU with 8 cores.

\begin{table}[!h]
\centering
% \tiny
\vspace{-0.5cm}
\caption{The 19 families under Class \textit{Caudoviricetes} from the RefSeq database we used in the experiments. Number: the number of complete sequences in each family. }
\renewcommand\arraystretch{1.2}
\begin{tabular}{ p{3.6cm}p{3.6cm}p{3.6cm}p{3.6cm} }
\hline
\textbf{Family Name} & \textbf{Number} & \textbf{Family Name} & \textbf{Number} \\ \hline
\textit{Autographiviridae} & 370 & \textit{Straboviridae} & 204 \\
\textit{Herelleviridae} & 127 & \textit{Drexlerviridae} & 117\\
\textit{Demerecviridae}  & 94 & \textit{Peduoviridae} & 83\\
\textit{Casjensviridae}    & 76 & \textit{Schitoviridae} & 76 \\
\textit{Kyanoviridae}  & 62 & \textit{Ackermannviridae}  & 62  \\
\textit{Rountreeviridae}   & 35 & \textit{Salasmaviridae}  & 34\\
\textit{Vilmaviridae} & 31 & \textit{Zierdtviridae} & 26 \\
\textit{Mesyanzhinovviridae} & 17 & \textit{Chaseviridae} & 14\\
\textit{Zobellviridae}  & 13 & \textit{Orlajensenviridae} & 11 \\
\textit{Guelinviridae}  & 8 &\textit{\textbf{Total}}& 1460\\
\hline
\end{tabular} 
\label{tab:family}
\end{table}

\subsection{Dataset}

We rigorously evaluated these phages taxonomic classification tools on multiple datasets. The detailed information is listed below.

\begin{itemize}
\item[$\bullet$] \textit{The RefSeq dataset} RefSeq is a widely used benchmark dataset in phage classification tasks. By October 2022, there are 1,826 complete sequences with family-label under Class \textit{Caudoviricetes} in the RefSeq database. In this paper, we only focus on the phages infecting bacteria. After filtering out the families that infect archaeas or contain sequences less than 6, there are 19 families (including 1460 complete sequences) we can use in our experiments. Table \ref{tab:family} shows the number of sequences within the 19 families under class \textit{Caudoviricetes}, among which \textit{Autographiviridae} contains the largest number of sequences. For the tools that require protein sequences, we used Prodigal \cite{hyatt2010prodigal} to predict and translate the nucleotide sequence into the proteins.

We sorted the sequence by their release time at RefSeq. Then, we used the first  80\% of the labeled complete sequences from each family as the training set/reference database to retrain/update the four tools, and the rest 20\% as test set. Because we split the data in chronological order, the data in the test set are more recent (almost all were released in 2020 or after).

\item[$\bullet$] \textit{Short contigs dataset} This dataset contains segments with different lengths, including 500 bp, 1,000 bp, 3,000 bp, 5,000 bp, 10,000 bp, and 15,000 bp. We randomly generated the segments from the 20\% RefSeq dataset (293 sequences) mentioned above. For each length, we cut ten segments from each phage genome by selecting a random start position. Finally, we had 2, 930 phage contigs for each length and 29,300 for all different lengths. Then, we used these segments to evaluate the performance of the four tools on short contigs.

\item[$\bullet$] \textit{Simulated metagenomic dataset} We used a simulated metagenomic dataset generated by six common bacteria living in human gut \cite{Shang_2022}. We first utilized metaSPAdes \cite{nurk2017metaspades} to assemble the reads into contigs. Then PhaMer \cite{Shang_2022} was applied to identify bacteriophages from metagenomic data, and the labels of the contigs were determined using BLAST \cite{camacho2009blast+}. Eventually, 37 contigs were used in the experiments. More details about this dataset will be provided in the section of \textbf{Experiment 4}.

\item[$\bullet$] \textit{Low-similarity dataset} To test the tools' performance on classifying highly diverged phages, we constructed a hard case where the test sequences share low similarity with the reference database/training data. Specifically, we calculated the Dashing pairwise similarity of the sequences in each family and then used the approach in \cite{petti2022constructing} to partition the data into two parts with specified maximum similarity. With this method, we got 264 and 45 genomes for training and test, where each test genome has at most 0.015 Dashing similarity with any reference genome. Then we randomly cut 15 contigs with a length of 3,000 bp and 5,000 bp, respectively, from each testing genome. Finally, there are 675 contigs for each length in the test set.
\end{itemize}

\subsection{Evaluating criteria for different tools}

\subsubsection{\textbf{Metrics}} \label{para:metrics} An ideal phage classification tool should assign correct labels for as many inputs as possible. Nevertheless, there is usually a tradeoff between the percentage of prediction and the accuracy of the prediction. Some tools may sacrifice the percentage of prediction in order to achieve high specificity and accuracy, while others may predict more with lower accuracy. Thus the first metric is \textit{prediction rate}, which is the ratio of outputs with prediction results ($N_{pred}$ in Equation \ref{equa:pred_rate}) to the total input ($N_{all}$ in Equation \ref{equa:pred_rate}). Because some tools only provide a family name as output, commonly used metrics such as AUROC cannot be computed. In this work, we calculated accuracy, recall, and precision for each tool (Equations \ref{equa:acc}-\ref{equa:recall}). $N_{correct}$ is the number of sequences with correct predictions in output.  $N_{total}$ is the total number of sequences used to evaluate, which can be $N_{all}$ or $N_{pred}$ when we report accuracy for all input phage sequences ($N_{all}$) or only for sequences with predictions in output ($N_{pred}$), respectively. Providing accuracy for all input sequences has the advantage of using the same denominator (i.e. $N_{all}$) for all tools. But it penalizes the tools of low prediction rate twice. On the other hand, reporting accuracy for only sequences with predictions removes the impact of prediction rate but may favor tools with low prediction rate (i.e. small $N_{pred}$). Thus, reporting both can provide a more comprehensive evaluation for users. For example, if there are 293 ($N_{all}$) sequences input, among which 290 sequences have classification prediction results ($N_{pred}$), and 285 of them have correct results ($N_{correct}$), the accuracy on all input will be 285/293=0.973, and the accuracy on predicted sequences will be 285/290=0.983. We only calculate the recall and precision of each family ($Precision_i$ and $Recall_i$) to check the performance on different families. \textit{$TP_i$}, \textit{$FP_i$}, and \textit{$FN_i$} are the true positive, false positive, and false negative for family $i$, respectively.

\begin{equation}
Prediction\ rate = \dfrac{N_{pred}}{N_{all}}
\label{equa:pred_rate}
\end{equation}

\begin{equation}
Accuracy = \dfrac{N_{correct}}{N_{total}}
\label{equa:acc}
\end{equation}

\begin{equation}
{Precision_i} =\dfrac{TP_i}{TP_i+FP_i}
\label{equa:precision}
\end{equation}

\begin{equation}
{Recall_i} = \dfrac{TP_i}{TP_i+FN_i}
\label{equa:recall}
\end{equation}

\subsubsection{\textbf{Description of the output}} 
Because the output format of each tool is different, we will describe how we process the output and calculate the metrics in detail.

vConTACT 2.0 can output the result of each sequence and assign it a ``VC State'', including ``Singleton'', `Outlier'', or ``Clustered''. In addition, the sequences with a ``Clustered'' state will be assigned to a VC cluster/subcluster. When the query sequence is within the same VC cluster as a reference genome, the taxonomic labels can be assigned based on the known labels. However, some sequences are clustered but have no reference genome in the same VC cluster, so they can not be assigned with a known label. Therefore, we treat the sequence with VC state of ``Singleton'', ``Outlier'', and ``Clustered'' but no reference genome in the same clusters, as ``no prediction''. In other words, $N_{pred}$ of vConTACT 2.0 refers to the number of the sequences that are clustered with reference genomes.

PhaGCN will not output the classification results for the sequences they can not classify, so $N_{pred}$ of PhaGCN is the number of sequences that can be predicted.

MMseqs2 and CAT will not output any prediction result for the sequences they cannot classify. The classification result of MMseqs2 and CAT can be a label at different ranks. If the prediction at the lowest rank is above family, we also treat this sequence as ``no prediction'' for the family level. The number of the rest sequences is $N_{pred}$ of MMseqs2/CAT.

\subsection{Experiment 1: Leave-one-family-out experiments} 
The constant change of ICTV underscores a need for classification tools to recognize the sequences that are not part of the current classification system. For example, the three largest families, \textit{Siphoviridae}, \textit{Podoviridae}, and \textit{Myoviridae},  were largely removed from the current ICTV system. Some of the sequences that belonged to these three families are not part of any existing family. Thus, the classification tools need to handle these out-of-distribution sequences by providing a signal for users.

To examine whether the tested tools can single out those out-of-distribution sequences, we removed all the phages in one family from the training data and retrained the models. Then the retrained models are applied to the removed family members. Ideally, the test sequences in this removed family should not be classified into any existing family labels.

\begin{figure}[h!]
    \centering
    \vspace{-0.2cm}
    \hspace{-0.8cm}
    \setlength{\belowdisplayskip}{0pt}
    \includegraphics[width = 0.9\linewidth]{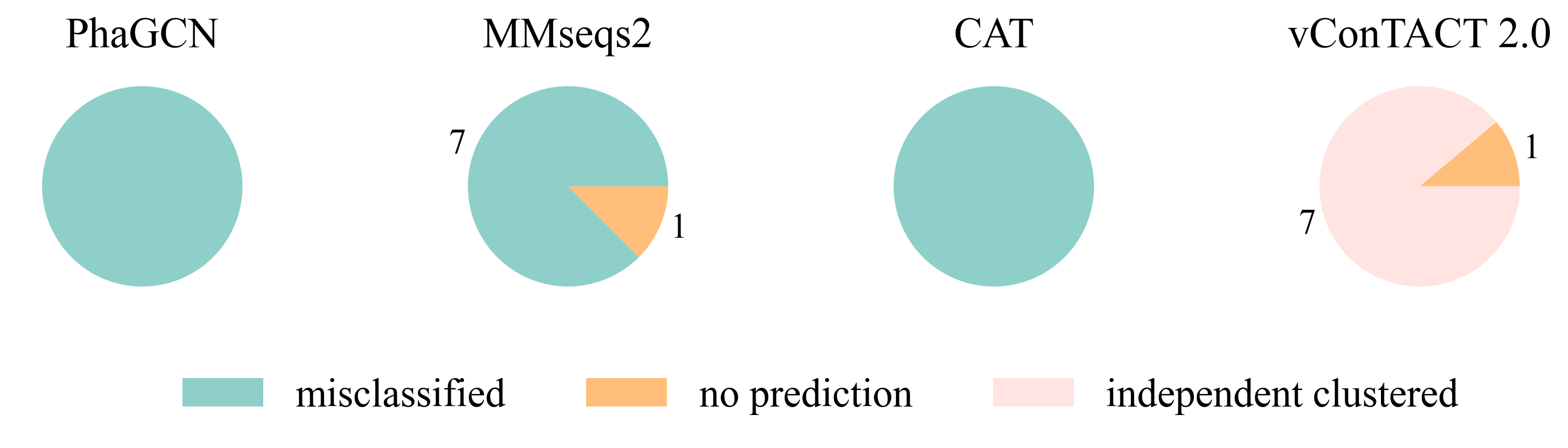}
    \caption{The classification result of \textit{Guelinviridae} sequences in tools that are retrained by removing all \textit{Guelinviridae} sequences. ``independent clustered'': the sequences are in a VC cluster without any reference genome. }
    \label{fig:gue-pie}
\end{figure}

\begin{figure}[h!]
    \centering
    \hspace{-0.8cm}
    \setlength{\belowdisplayskip}{0pt}
    \includegraphics[width = 0.9\linewidth]{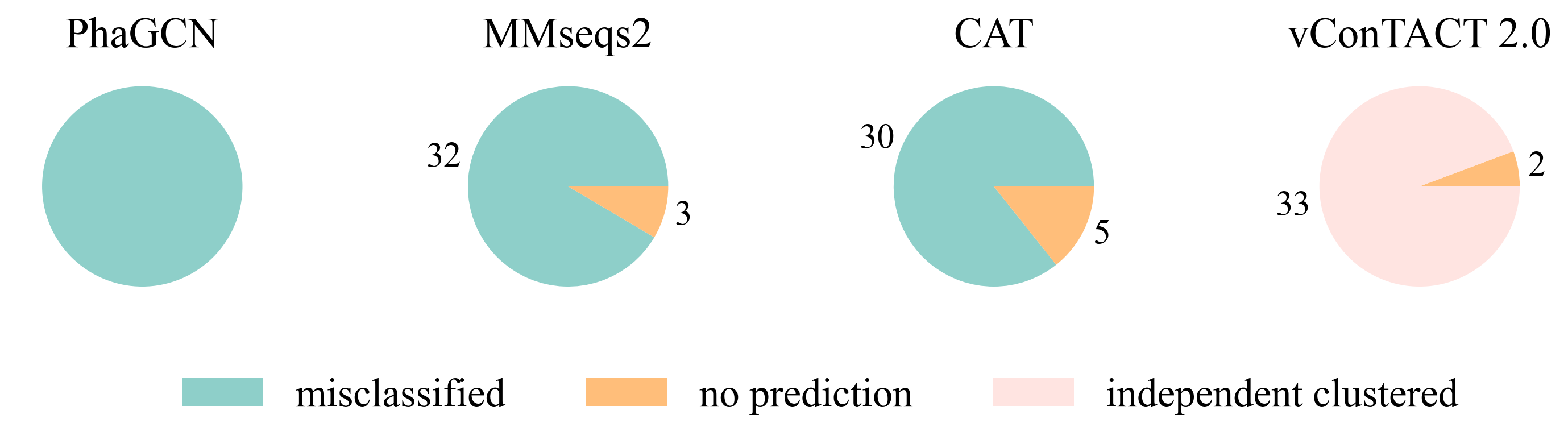}
    \caption{The classification result of \textit{Rountreeviridae} sequences in tools that are retrained by removing all \textit{Rountreeviridae} sequences. ``independent clustered'': the sequences are in a VC cluster without any reference genome.}
    \label{fig:roun-pie}
\end{figure}

At first, we conducted the experiments on a small and a relatively large family: \textit{Guelinviridae} and \textit{Rountreeviridae}. The classification results are plotted in Fig. \ref{fig:gue-pie} and Fig. \ref{fig:roun-pie}, which show that PhaGCN assigned all of the query genomes to one of the other families in the training set, while CAT and MMseqs2 can correctly recognize a few sequences as ``no family label''. However, vConTACT 2.0 can assign all sequences to ``Outlier/Singleton''  or a ``VC cluster'' without reference genomes.

\begin{table}[!h]
\centering
% \tiny
\caption{The percentage of misclassified sequences in leave-one-family-out experiment for each family.}
\renewcommand\arraystretch{1.2}
\begin{tabular}{ p{3.5cm}p{3.5cm}p{3.5cm}p{3.5cm} }
\hline
\textbf{Family Name} & \textbf{CAT} & \textbf{MMseqs2} & \textbf{vConTACT 2.0} \\ \hline
\textit{Autographiviridae} & 0.78 & 0.58 & 0.12 \\
\textit{Straboviridae} & 0.05 & 0 & 0 \\
\textit{Herelleviridae} & 0.02 & 0.2 & 0\\
\textit{Drexlerviridae} & 0.3 & 0.25 & 0.01\\
\textit{Demerecviridae}  & 0 & 0.32 & 0\\
\textit{Peduoviridae} & 0.55 & 0.84 & 0\\
\textit{Casjensviridae} & 0.86 & 0.87 & 0 \\
\textit{Schitoviridae} & 0.24 & 0.21 & 0 \\
\textit{Kyanoviridae}  & 0 & 0.18 & 0 \\
\textit{Ackermannviridae}  & 0 & 0.02 & 0 \\
\textit{Rountreeviridae}   & 0.86 & 0.91 & 0 \\
\textit{Salasmaviridae}  & 0.88 & 0.82 & 0.44 \\
\textit{Vilmaviridae} & 0.13 & 0.45 & 0 \\
\textit{Zierdtviridae} & 0.81 & 0.92 &0 \\
\textit{Mesyanzhinovviridae} & 0 & 0.06 & 0 \\
\textit{Chaseviridae} & 0 & 0.14 & 0\\
\textit{Zobellviridae}  & 0.77 & 0.38 & 0 \\
\textit{Orlajensenviridae} & 0.73 & 0.18 & 0 \\
\textit{Guelinviridae}  & 1.0 & 0.88 & 0 \\
\hline
\textbf{Average} & 0.42 & 0.43 & 0.03\\
\hline
\end{tabular} 
\label{tab:leave-family}
\end{table}

We then extended the experiment to each family. Because the current version of PhaGCN is not designed to handle out-of-distribution sequences, we only show the results for CAT, MMseqs2, and vConTACT 2.0 in Table \ref{tab:leave-family}. The output of these three tools for the test sequences are divided into two parts: those that did not output a family label (``no prediction'', defined in the section \textbf{Description of the output}), and those that can output a family label from the training data (i.e., a misclassification in this experiment). Table \ref{tab:leave-family} shows the misclassification rate of each tool. CAT and MMseqs2 assign more test sequences to other families in the reference database. In contrast, vConTACT 2.0 can assign almost all sequences of each family to  ``Outlier/Singleton'' labels or ``VC cluster'' without reference genomes. The misclassification rates of CAT and MMseqs2 vary widely across different families, with the ranges 0-1 and 0-0.92, respectively. A closer look at those results reveals that the misclassified phages tend to distribute in a small set of families. For example, almost all sequences belonging to \textit{Guelinviridae} are classified into \textit{Salasmaviridae} by CAT, which is likely due to the higher inter-family similarity between them. Specifically, 29.6\% proteins of \textit{Guelinviridae} can align with \textit{Salasmaviridae} using BLASTP. Similarly, sequences from \textit{Zobellviridae} tend to be classified into family \textit{Autographiviridae} because they share about 16.9\% proteins. Therefore, the inter-family similarity is an essential factor leading to misclassification. Overall, the misclassification results of MMseqs2 are more divergent than CAT. For example, CAT will classify \textit{Autographiviridae} genomes into 4 other families, while MMseqs2 will assign them into 8 families (including the 4 families in CAT).

\begin{figure}[h!]
    \centering
    \hspace{-0.8cm}
    \setlength{\belowdisplayskip}{0pt}
    \includegraphics[width=0.9\linewidth]{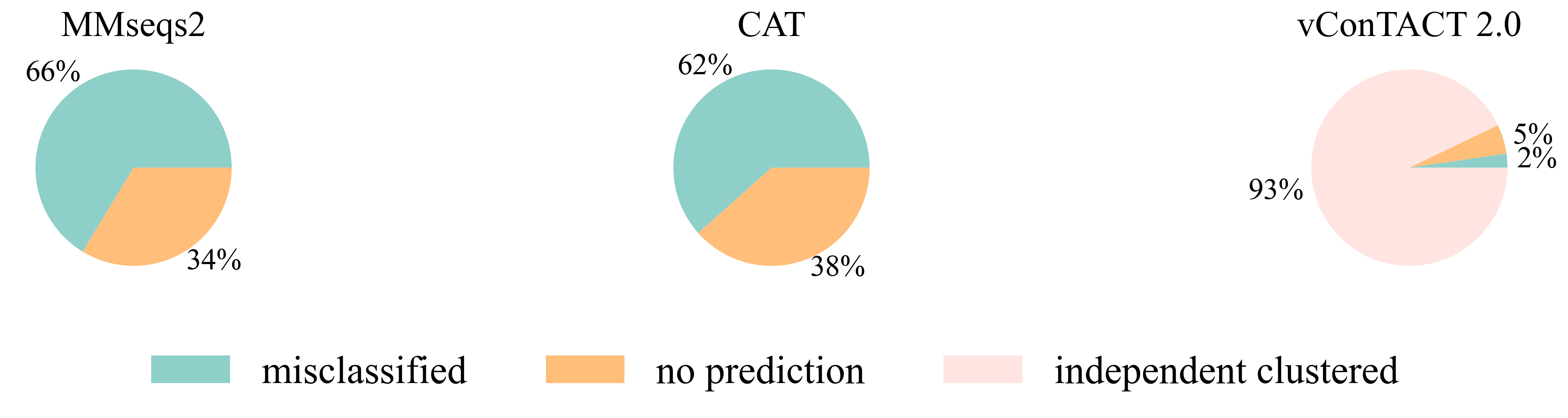}
    \caption{The classification result of 2445 unclassified sequences. ``independent clustered'': the sequences are in a VC cluster without any reference genome.}
    \label{fig:df2450-pie}
\end{figure}

Then we extended the experiment to the genomes that are unclassified at the family level in the RefSeq database under class \textit{Caudoviricetes}. Because the three largest families \textit{Myoviridae}, \textit{Siphoviridae} and \textit{Podoviridae} were removed, we used the genome sequences that initially belonged to these three families but now no longer have a family label as the test data. There are 2445 of them, and the classification result is shown in Fig. \ref{fig:df2450-pie}. 
MMseqs2 and CAT misclassified about 65\% of the input sequences. vConTACT 2.0 can identify 98\% unclassified sequences by assigning them in independent clusters or outputting a ``Singleton/Outlier'' label and only misclassified 2\% sequences.
In conclusion, vConTACT 2.0 performs better in identifying novel phages than the other three tools.

\subsection{Experiment 2: classification performance} 
As we described in Section \textbf{``Dataset''}, we used 20\% (293) of the complete sequences from the RefSeq database as the test set, and the other 80\% as the reference/training set. To mimic metagenomic assembled contigs, we generated six sets of segments of different lengths for comparison, including 500 bp, 1,000 bp, 3,000 bp, 5,000 bp, 10,000 bp, and 15,000 bp. We randomly selected the start positions for each length and cut ten segments from each complete sequence. Finally, we had 2,930 phage fragments for each length and 29,593 for all different lengths as the test data (293 complete sequences + 2930 * 10 short fragments).

\begin{figure}[h!]
    \centering
    \hspace{0.4cm}
    \includegraphics[width=1\linewidth]{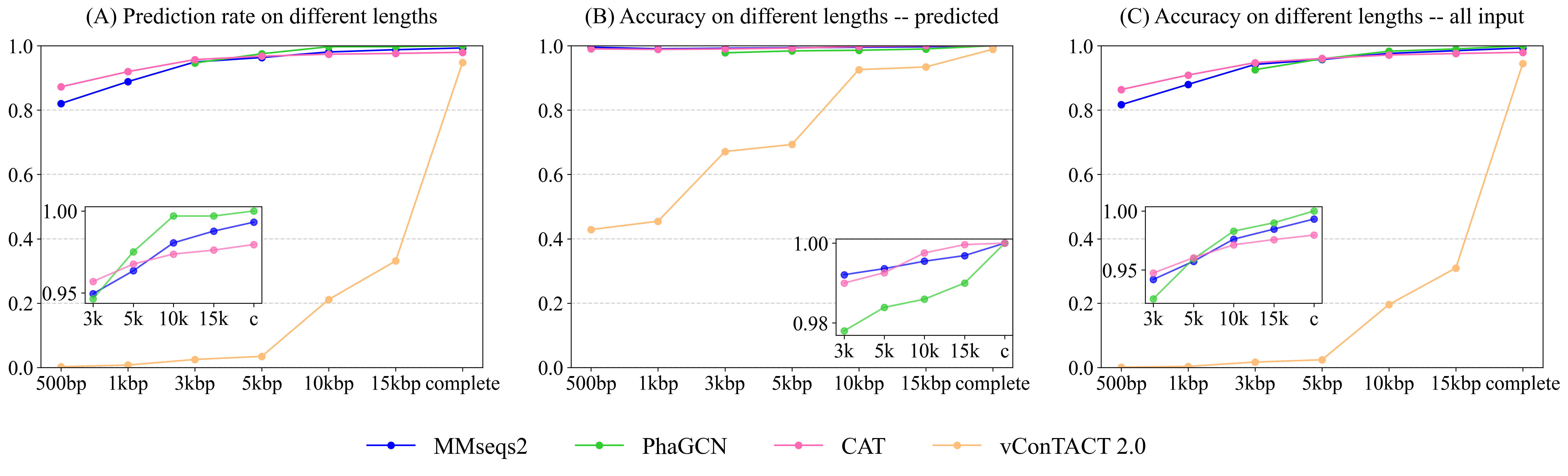}
    \vspace{-0.6cm}
    \caption{The performance of each tool on contigs from the RefSeq. (A): the prediction rate of four tools on different lengths. (B): the accuracy of four tools on phage contigs with predictions. (C): the accuracy of four tools on all input phage contigs. X-axis: the lengths Y-axis: the values.}
    \label{fig:refseq}
\end{figure}

A good taxonomic classification tool should have a high prediction rate and high accuracy. First, we recorded the prediction rate of each tool on different lengths. Because PhaGCN only accepts contigs longer than 2,000 bp, we do not show its results on 500 bp and 1,000 bp in Fig. \ref{fig:refseq}. The prediction rate (Fig. \ref{fig:refseq} (A)) of all tools becomes higher with the increase in sequence length. This is expected because longer sequences usually provide more information for classification. Almost all pipelines can maintain a high prediction rate (\textgreater 80\%) on short sequences except vConTACT 2.0. PhaGCN has the highest prediction rate if the inputs are longer than 5,000 bp, while CAT is slightly lower. vConTACT 2.0 is mainly designed for complete or long sequences, and its prediction rate drops sharply when the inputs are shorter than 15,000 bp. All four can handle more than 95\% of complete sequences, among which PhaGCN can predict all of them (100\%), and the prediction rates of MMseqs2, CAT, and vConTACT 2.0 are 99.3\%, 97.9\%, and 95.1\%, respectively.

Fig. \ref{fig:refseq} (B) shows the accuracy of the four tools on phage sequences with predictions ($N_{pred}$ in Equation \ref{equa:pred_rate}). Similar to the prediction rates above, the accuracy of these approaches becomes better as the sequence lengths increase. The classification ability of CAT, PhaGCN, and MMseqs2 are not significantly affected by the change of contig lengths. On incomplete contigs, the accuracy of vConTACT 2.0 has an obvious upward trend when length increases. CAT gains the best prediction accuracy for contigs longer than 5,000 bp. Combined with the slightly lower prediction rate of CAT mentioned above, we can conclude that there is a tradeoff between the prediction rate and the accuracy of CAT. The accuracy of PhaGCN is slightly lower than the other two on contigs, and all three tools reach a high accuracy (100\%) for all complete sequences with predictions.

Fig. \ref{fig:refseq} (C) shows the accuracy of the four tools on all input phage contigs ($N_{all}$ in Equation \ref{equa:pred_rate}), which combines the results in (A) and (B) in order to display the overall performance of each tool. It reveals that PhaGCN keeps the best performance on contigs longer than 5,000 bp and reaches 100\% accuracy on complete genomes because it gains 100\% accuracy and prediction rate in (A) and (B), respectively. It is worth noting that the other three tools all have a less than 100\% recall on \textit{Autographiviridae}, most likely due to the lower pairwise similarity in \textit{Autographiviridae} (Table \ref{tab:similarity}). Due to the length limitation of PhaGCN, it is not suitable for classifying contigs shorter than 2,000 bp. When classifying contigs longer than 2,000 bp, PhaGCN and MMseqs2 are recommended for obtaining high prediction rates. Otherwise, CAT is a better choice if precision is the primary consideration.

\begin{figure}[h!]
    \includegraphics[width=1\linewidth]{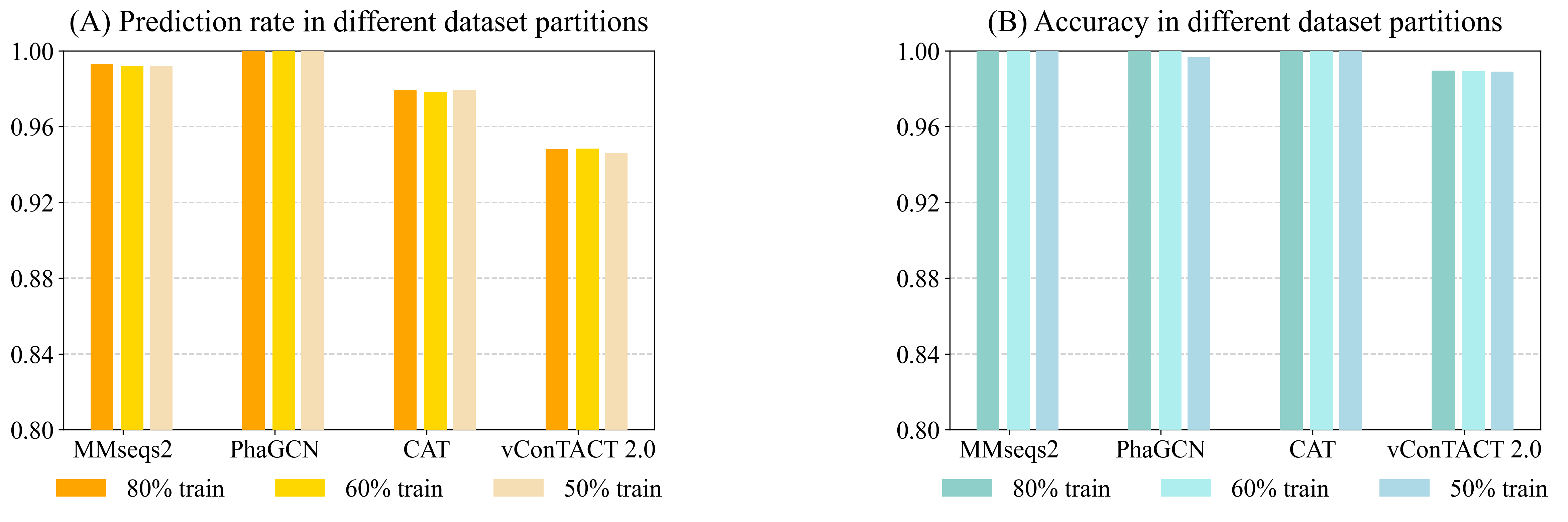}
    \vspace{-0.4cm}
    \caption{(A) The prediction rate of four tools with reduced reference datasets. (B) The corresponding accuracy on sequences with predictions. X-axis: the tools and training data partitions Y-axis: the values.}
    \label{fig:different-training}
\end{figure}

\subsection{Experiment 3: impact of training set size on classification performance}

Being a learning-based classification tool, PhaGCN can be affected by training data size. To test whether PhaGCN and other alignment-based tools suffer from reduced training data/reference database,  we used 80\% (the same as \textbf{Experiment 2}), 60\%, and 50\% of the RefSeq databases as the reference database for these tools, respectively. Then we tested them on the same test set as in \textbf{Experiment 2}. As shown in Fig. \ref{fig:different-training} (A), the prediction rates of PhaGCN with different reference databases have no obvious differences. There is a slight change in the prediction rate of CAT, MMseqs2, and vConTACT 2.0, but the differences do not exceed 0.2\%. In addition, the accuracy of these tools shown in Fig. \ref{fig:different-training} (B) are almost identical and are less affected than the prediction rate.

\begin{figure}[h!]
    \centering
    \setlength{\belowdisplayskip}{0pt}
    \includegraphics[width=1\linewidth]{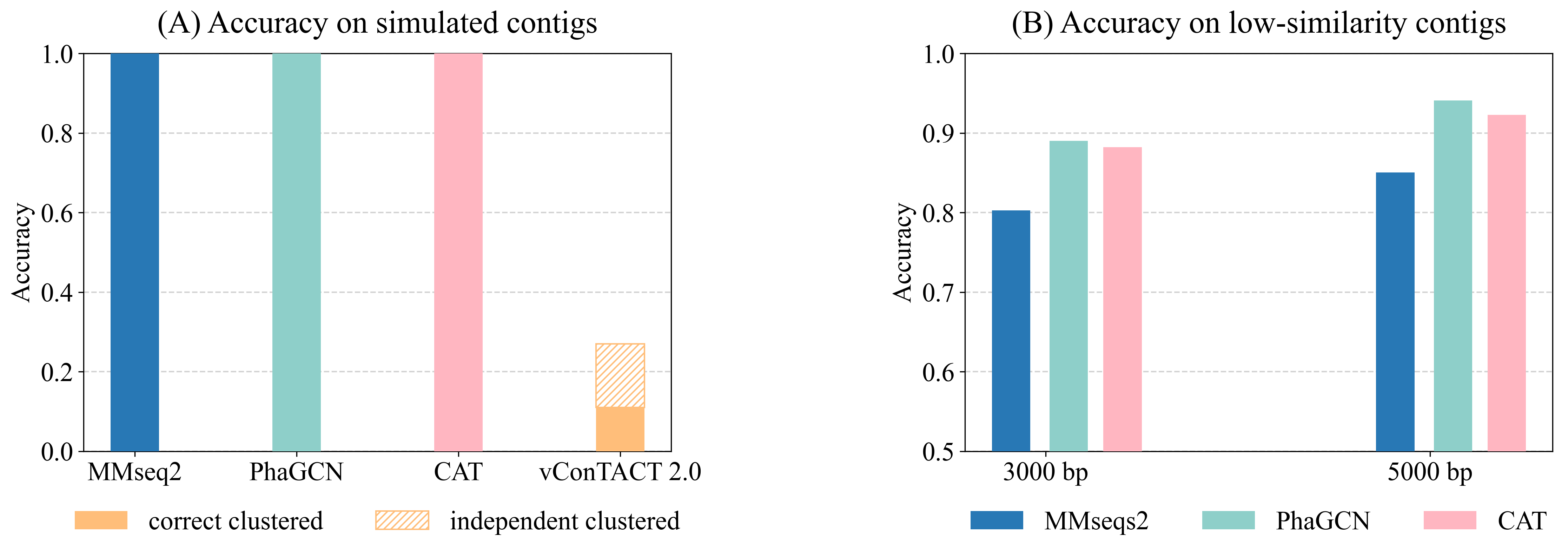}
    \caption{(A) The performance of the four tools on the simulated metagenomic dataset. The bars show the accuracy on all inputs. The top part with patterns in vConTACT 2.0 shows the percentage of contigs that are not clustered with any reference genome. (B) The performance of each tool on the two low-similarity datasets. Each bar shows the tools' accuracy on all input contigs.}
    \label{fig:sim-bar}
\end{figure}

\subsection{Experiment 4: classification performance on the simulated metagenomic dataset}

In this experiment, we used the simulated metagenomic dataset provided in PhaMer \cite{Shang_2022}. The dataset is a small-scale metagenomic dataset simulated by CAMISIM \cite{fritz2019camisim} using the commonly seen bacteria living in the human gut and the phages that infect these bacteria. The reads were assembled into contigs using metaSPAdes \cite{nurk2017metaspades}. 

\begin{table}[!h]
    \centering
    \renewcommand\arraystretch{1.3}
    \caption{Family composition of the simulated metagenomic dataset.}
    \begin{tabular}{p{8.2cm}p{7cm} }
    \hline
    \textbf{Family Name}       & \textbf{Number} \\ \hline
    \textit{Straboviridae} & 28\\
    \textit{Drexlerviridae} & 6\\ 
    \textit{Demerecviridae}  & 1 \\
    \textit{Peduoviridae} & 1 \\
    \textit{Ackermannviridae}  & 1  \\
    \textit{Total}     & 37 \\
    \hline
    \end{tabular} 
    \label{tab:sim_family}
\end{table}

We kept contigs of size above 3,000 bp. To assign labels to the contigs, we used BLAST \cite{camacho2009blast+} to map contigs to reference genomes and calculated the coverage. Only the contigs with at least 90\% of the sequence aligning to a reference genome were kept. Others are likely chimeric contigs due to assembly errors and thus are not used for testing. Finally, the number of contigs we could use in the experiment is 37. The name of the families and the number of genomes within each family are listed in Table \ref{tab:sim_family}. Compared to Table \ref{tab:family}, this test set contains a different abundance distribution for the component families, which can thus change the performance of these tools.

As shown in Fig. \ref{fig:sim-bar} (A), PhaGCN, MMseqs2, and CAT can classify all the simulated sequences correctly, which is slightly higher than that on the RefSeq data in \textbf{Experiment 2}. A plausible reason is that most of the sequences in this simulated dataset belong to \textit{Straboviridae} and \textit{Ackermannviridae}, which make up a large part of the reference database according to Table \ref{tab:family} (14\% and 4\%). In addition, they have greater intra-family similarities. 
The performance of vConTACT 2.0 is lower than the other three tools because the assembled contigs are short. This experiment shows that PhaGCN, MMSeq2, and CAT can process assembled contigs with different lengths.

\subsection{Experiment 5: classification performance on the low-similarity dataset}
Although the updated families under the new ICTV standard exhibit higher pairwise sequence similarity, there are still some diverged members. The diverged members may appear more often when sequencing new or underrepresented ecosystems. Thus, we test these tools' performance on predicting highly diverged sequences using the ``low similarity dataset''. There are 45 genomes in the test set with the maximum Dashing similarity of 0.015 with any reference genome. Then we randomly cut 15 contigs with a length of 3,000 bp and 5,000 bp from each test genome, leading to 1,350 contigs in total. Fig. \ref{fig:sim-bar} (B) shows the accuracy of all inputs. Because vConTACT 2.0 can not handle short contigs, we exclude it from this experiment.

Fig. \ref{fig:sim-bar} (B) reveals that the accuracy of MMseqs2 decreases by more than 10\% compared to  Fig. \ref{fig:refseq} (C) from \textbf{Experiment 2}. And the accuracy drop in CAT (6\%, 5.2\%) are greater than PhaGCN (3.3\%, 2\%) on the contigs of the same lengths. Therefore, the increased divergence between test and training data has a greater impact on alignment-based tools than PhaGCN in this experiment.

\subsection{Comparison of Running Time}

Running time is also an essential factor to consider for practical usage. Table 5 shows the running time of the tools for processing 500 complete sequences in RefSeq when using a different number of CPUs. Users can save more time by increasing the number of CPUs. The table also shows that CAT and MMseqs2 take the least time to process 500 complete phages.

\begin{table*}[!h]
\renewcommand\arraystretch{1.5}
{
\caption{The total running time of tools for classifying 500 genomes using a different number of CPUs. All the tools are run on IntelVR$^\circledR $XeonVR$^\circledR$  Gold 6258 R CPU with 8 cores.}
\begin{tabular}{p{2.7cm}|p{2.7cm}p{2.7cm}p{2.7cm}p{3.3cm}}
\hline
\textbf{Time (min)}& \textbf{PhaGCN} & \textbf{MMseqs2} & \textbf{CAT} & \textbf{vConTACT 2.0} \\ 
\hline
1 CPU  & 23 & 2 & 3 & 141\\
4 CPUs  & 18 & 1 & 2 & 64\\
\hline
\end{tabular}}
\label{tab:running-time}
\end{table*}

\section{Discussion and conclusion}
This work presents a review of taxonomic classification tools on phage family classification under \textit{Caudoviricetes}. To our best knowledge, this is the first review under the new ICTV standard released in August 2022. Compared to the previous version of ICTV, the updated families in the latest system are more conserved, which warrants a high prediction rate and accuracy of alignment-based tools. For example, the prediction rate of CAT and vConTACT 2.0 were 62\% and 92\% on the data in the previous ICTV system, respectively. And their accuracy on complete genomes were only 61.7\% and 86\%. However, their prediction rate and accuracy are significantly better under the new classification system.

%change update/retraining
The constant change of the taxonomic classification system by ICTV emphasizes the need for a tool to provide database updating or model retraining. Tools without these utilities can return obsolete or even wrong labels, making their practical usage limited. Many of these tools in Table \ref{tab:tool-compare} either lack this option or need excessive efforts to retrain.

%classification system is not complete
Despite great efforts, the current classification system by ICTV is not complete. New families can appear with new viruses sequenced and discovered, particularly those from underrepresented ecosystems. Thus, it is desired that a classification tool can handle out-of-distribution inputs, which are not part of any existing families. Based on our leave-one-family-out experiment, vConTACT 2.0 is more sensitive to those out-of-distribution sequences than others. 
However, a price paid by vConTACT 2.0 is its low prediction rate on short contigs, which is likely caused by the low gene sharing significance score between the query and the reference. Other tools perform better on short contigs, which is important for virus composition analysis in metagenomic data.

PhaGCN can only classify sequences on the family level. The lowest levels that the other three tools can classify are genus level or below. The experimental results show that all of them can perform well on complete genomes from the RefSeq database after retraining. PhaGCN has the highest prediction rate when classifying short contigs (\textgreater 3,000 bp), and CAT gains a higher accuracy with a slightly lower prediction rate. Therefore, when classifying incomplete contigs larger than 3,000 bp, PhaGCN, CAT, and MMseqs2 can all be considered, but PhaGCN has a better overall performance. In addition, CAT and MMseqs2 can be used to classify contigs shorter than 2,000 bp because PhaGCN can not handle that length. 
All these four tools are robust against the size reduction of the reference database/training data. The performance of PhaGCN is less affected in classifying highly diverged sequences that share low similarity with the reference genomes.

The focus of this review is family-level classification. While the current families annotated by ICTV usually contain multiple phages per family, the genus size distribution exhibits a much more skewed distribution with many genera only containing one phage genome. It is not trivial to create appropriate reference database/training data vs. test data with hundreds of rare genera. It is our future work to examine the impact of the long tail distribution on current classification tools.

\section*{Data Availability}
The detailed information of the code and datasets is provided in the supplementary file.

\section*{Funding}
City University of Hong Kong (Project 9678241 and 7005453) and the Hong Kong Innovation and Technology Commission (InnoHK Project CIMDA).

\bibliographystyle{unsrt}  
\bibliography{references}  

\end{document}